\documentclass{aa}  

\usepackage{natbib,twoopt}
\usepackage{multirow}
\usepackage{amsmath}
\usepackage{amssymb} 
\usepackage[T1]{fontenc}
\usepackage{blindtext}
\usepackage{hyperref}
\hypersetup{
    colorlinks=true,
    linkcolor=red,
    filecolor=blue,
    citecolor=blue,
    urlcolor=blue,
    pdftitle={Overleaf Example},
    pdfpagemode=FullScreen,
}
\usepackage{graphicx}
\usepackage{gensymb}
\usepackage{subcaption}
\usepackage{pythonhighlight}
\usepackage{listings}
\usepackage{xcolor}
\usepackage{multirow}
\usepackage{siunitx}
\usepackage{makecell}
\usepackage{orcidlink}
\usepackage{float}

\usepackage{amsmath} 

\usepackage[varg]{txfonts}

\usepackage{graphicx}	
\usepackage{amsmath}	

\definecolor{codegreen}{rgb}{0,0.6,0}
\definecolor{codegray}{rgb}{0.5,0.5,0.5}
\definecolor{codepurple}{rgb}{0.58,0,0.82}
\definecolor{backcolour}{rgb}{0.95,0.95,0.92}

\lstdefinestyle{mystyle}{
  backgroundcolor=\color{backcolour}, commentstyle=\color{codegreen},
  keywordstyle=\color{magenta},
  stringstyle=\color{codepurple},
  basicstyle=\ttfamily\footnotesize,
  breakatwhitespace=false,         
  breaklines=true,                 
  captionpos=b,                    
  keepspaces=true,                                               
  showspaces=false,                
  showstringspaces=false,
  showtabs=false,                  
  tabsize=2
}

\begin{document}

\title{Accurate distances of the Galactic spiral arms from dust-scattered X-ray emission of gamma-ray bursts}

\author{B. Vaia\inst{1,2,3} \orcidlink{0000-0003-0852-0257}, I. Fornasiero\inst{1,2,3} \orcidlink{0009-0006-0909-7373}, A. Tiengo\inst{1,3}\orcidlink{0000-0002-6038-1090}, A. Bracco \inst{4,5}\orcidlink{0000-0003-0932-3140}, V. Jeli\'c\inst{6}\orcidlink{0000-0002-6034-8610}, \v{Z}. Bo\v{s}njak\inst{7}\orcidlink{0000-0001-6536-0320}, F. Pintore\inst{8}\orcidlink{0000-0002-3869-2925} \and  P. Esposito\inst{1,3}\orcidlink{0000-0003-4849-5092}}
          
\authorrunning{B. Vaia et al.}

\institute{Scuola Universitaria Superiore IUSS Pavia, Piazza della Vittoria 15, 27100 Pavia, Italy\\ \email{beatrice.vaia@inaf.it}
  \and Department of Physics, University of Trento, Via Sommarive 14, 38123 Povo (TN), Italy 
  \and Istituto Nazionale di Astrofisica, Istituto di Astrofisica Spaziale e Fisica Cosmica di Milano, via A. Corti 12, 20133 Milano, Italy
  \and INAF - Osservatorio Astrofisico di Arcetri, Largo E. Fermi 5, 50125 Firenze, Italy
  \and Laboratoire de Physique de l'Ecole Normale Sup\'erieure, ENS, Universit\'e PSL, CNRS, Sorbonne Universit\'e, Universit\'e de Paris, F-75005 Paris, France
  \and Ru\dj er Bo\v{s}kovi\'c Institute, Bije\'nicka cesta 54, 10000 Zagreb, Croatia
  \and University of Zagreb Faculty of Electrical Engineering and Computing, Unska ul. 3, 10000 Zagreb, Croatia
  \and  INAF - Istituto di Astrofisica Spaziale e Fisica Cosmica di Palermo, Via U. La Malfa 153, 90146 Palermo, Italy}
   \date{Received 26/09/2025; accepted 27/04/2026}


\abstract{
The details of the spiral structure of the Milky Way are still debated due to large uncertainties in the distance estimates obtained through the most common tracers. X-ray dust scattering rings produced by short extragalactic X-ray transients provide instead a direct method to measure the 3D distribution of interstellar clouds up to the edges of our Galaxy with a few percent precision.

We report on the analysis of all the available \textit{XMM-Newton} and \textit{Chandra} follow-up observations of three low-latitude gamma-ray bursts: GRB 031203 ($l \simeq 255\degree$, $b \simeq -5\degree$), GRB 160623A ($l \simeq 84\degree$, $b \simeq -3\degree$), and GRB 221009A ($l \simeq 53\degree$, $b \simeq 4\degree$).
The previous detection of X-ray rings in these observations,
produced by dust clouds located beyond 5 kpc, can be associated with dust in the Perseus, Outer, and 
Outer Scutum-Centaurus arms, thus providing direct distance measurements to these 
structures along three distinct lines of sight.

We have identified two additional rings in the direction of GRB 160623A, produced by dusty clouds at $6.91\, \pm\,0.06$ kpc and $9.9\,\pm\,0.6$ kpc, and confirmed -- through a second \textit{XMM-Newton} observation -- the presence of one cloud at $9.7\,\pm\,0.4$ kpc toward GRB 031203. We also accurately measured the distance of dusty clouds up to $19.0\,\pm\,0.2$ kpc owing to the analysis of one \textit{Chandra} and four \textit{XMM-Newton} observations of GRB 221009A. The small statistical and systematic uncertainties of these measurements place tight constraints on the geometry of the outer Milky Way and reveal significant deviations from current models, which critically depend on spectroscopy-based Galactic rotation curves at large distances.
}

\keywords{}

\maketitle
\nolinenumbers
\section{Introduction}\label{sec1}

Despite extensive studies throughout the last century, the exact spiral structure of the Milky Way is not yet well defined.
The Galactic structure is reconstructed on the basis of different tracers, whose distance is often affected by large uncertainties arising from kinematic distance ambiguities and the dependence on a Galactic rotation curve model.

The most widely used tracers (e.g. \citealt{2005A&A...440..775K}, \citealt{2009MNRAS.392..783G}, \citealt{2014A&A...569A.125H}, \citealt{Reid2019}) are 6.7 GHz methanol masers, giant molecular clouds (GMCs), \ion{H}{II} regions, and clouds of neutral atomic hydrogen (\ion{H}{I}). 
\ion{H}{II} regions and methanol masers are predominantly associated with star-forming regions, and GMCs serve as the birthplaces of massive stars. Because of this, their spatial distribution is expected to be consistent with that of the Galactic young stellar population. 

Unlike visible light from stars, the radio recombination lines emitted by \ion{H}{II} regions and the spectral lines from methanol masers are largely unaffected by absorption in the interstellar medium (ISM). This makes them particularly useful for probing distant and obscured regions of the Milky Way (e.g., \citealt{2011ApJS..194...32A}, \citealt{2015MNRAS.450.4109B}).
When it comes to identifying molecular clouds, the most commonly used tracer is carbon monoxide (CO). CO is the second most abundant molecule in the ISM after molecular hydrogen ($\rm{H_2}$).  Directly observing $\rm{H_2}$, however, is highly challenging due to the lack of a permanent dipole moment.
For these reasons, our knowledge of the spiral structure of the Milky Way is primarily based on CO surveys such as the CfA-Chile \citep{Dame2001,Dame2022}, FCRAO Galactic Ring Survey and Outer Galaxy Survey \citep{Heyer2001,Roman2010}, and SEDIGISM \citep{Schuller2021}. 
Another key component for tracing the Galactic structure is \ion{H}{I}, the most abundant element in the ISM, observed through the 21-cm hyperfine transition line. Surveys like HIPASS \citep{Barnes2001}, GALFA-HI \citep{Peek2011}, GASS \citep{McClure2009,Kalberla2010,Kalberla2015} and EBHIS \citep{Winkel2016} have played a crucial role in mapping the distribution of the ISM. 
In parallel with tracer-based catalogs, the Galactic spiral structure has also been investigated through top-down reconstructions based on radial velocity measurements. In this approach, spectroscopic observations of extended \ion{H}{I} or CO emission are combined with an assumed Galactic rotation curve to convert line-of-sight velocities into Galactocentric positions, producing maps of the Milky Way disk. Classical examples are the \ion{H}{I}-based reconstructions of \citet{Levine2006} and the work of \citet{Miville2017}, derived from a hierarchical decomposition of CO emission over the entire Galactic plane.
The choice and accuracy of the rotation curve are therefore crucial in assigning kinematic distances of Galactic objects.
Many rotation curve models have been developed over the last decades (such as the ones of \citealt{Clemens1985}, \citealt{1996MNRAS.281...27P}, \citealt{2019ApJ...871..120E}, \citealt{Reid2019}), but significant uncertainties still persist mainly regarding the distribution of the dark-matter component in the outer regions.

In 2013, the launch of the ESA Gaia satellite \citep{GaiaCollab} revolutionized the study of the structure of the Milky Way by providing precise astrometric measurements of stars (e.g. \citealt{Sale2018,Chen2018,Green2019,Hottier2020,Razaei2020,Lallement2022}). However, stellar distances measured by Gaia are available only up to a few kpc.
To explore more distant regions of the Galactic plane, extinction maps based on infrared (IR) data can be utilized. Over the years, significant work has been done in this field. For instance, \citet{Schlegel1998} used far-infrared dust emission data from the IRAS and COBE satellites to create an all-sky dust column density map. In 2006, \citet{Lombardi2006} employed 2MASS photometric data and the color excess technique to study molecular clouds, while \citet{Schlafly2010} used data from the Sloan Digital Sky Survey (SDSS). \citet{Rezaei2024} produced a 3D map of the Milky Way's plane extending up to 10 kpc using near-IR data from the SDSS Apache Point Observatory Galactic Evolution Experiment (APOGEE).
Another three-dimensional dust map obtained by combining Gaia parallaxes with large photometric surveys has been recently published by \citet{zucker25}. However, this map is limited to a confined portion of the Galactic plane only observable from the Southern hemisphere and, depending on the sightlines, it probes distances up to only $\sim$5--10 kpc.

Recent studies of the Milky Way's  structure \citep{Hou2014} suggest that our Galaxy has four spiral arms, which contrasts with earlier models that indicated only two arms \citep{Oort1958}.
However, uncertainties persist regarding the exact morphology of the Milky Way.
For example, only in 2011  \citet{Dame2011} identified a new spiral arm beyond the Outer Arm at $\simeq 15$ kpc from the Galactic center, at longitudes from $13\degree$ to $55\degree$ in the first quadrant. This new arm was discovered by studying CO emission and it was overlooked in existing 21 cm surveys probably because it is highly inclined relative to the Galactic plane.

X-ray dust scattering rings observed around gamma-ray bursts (GRBs) provide a unique opportunity to measure the distances to Galactic interstellar clouds \citep{BarbaraPaper}. These rings are formed by the prompt X-ray emission of the gamma-ray burst, whose photons reach the observer with an increasingly larger delay and angular offset due to scattering within the intervening dust clouds.
The accuracy of this method is very good up to large distances, allowing us to tightly constrain the distances of the outermost spiral arms.
For an extragalactic X-ray source and a dust cloud within our Galaxy, the angular size of the observed dust-scattering ring is given by:

\begin{equation} \theta [\rm{arcmin}]= 4.455\sqrt{\frac{\Delta t[\rm{days}]}{D_{\rm{d}}[\rm{kpc}]}},
\label{eq:theta}
\end{equation}
where $\Delta t$ represents the time delay between the direct and scattered X-rays, and $D_{\rm{d}}$ is the distance to the dust cloud.

To constrain the structure of our Galaxy, particularly focusing on the most distant arms, we identified three low-latitude ($|b|<5\degree$) GRBs around which X-ray scattering rings were detected by \textit{XMM-Newton} or \textit{Chandra}: GRB 221009A ($l\simeq53\degree$, $b\simeq4\degree$; \citealt{Tiengo2023,Williams2023}), GRB 160623A ($l\simeq84\degree$, $b\simeq-3\degree$; \citealt{Pintore2017}), and GRB 031203 ($l\simeq255\degree$, $b\simeq-5\degree$; \citealt{Vaughan2004,Tienghetti2006,Feng2010}).
These measurements can be used to constrain the structure of the Perseus, Outer, and Outer Scutum-Centaurus (OSC) arms.
The paper is structured as follows: Section 2 describes the data reduction process, Section 3 explains the analysis methodology and reports the results, which are discussed in Section 4, and Section 5 presents the conclusions.

\section{Observations and data reduction}\label{sec2}
In this work we report the analysis of X-ray data collected by the \textit{XMM-Newton} \citep{citeXMM} and \textit{Chandra} \citep{2000SPIE.4012....2W}  missions (Obs IDs in Table \ref{tableobsid}). The study of X-ray expanding rings requires a detailed characterization of the spatial distribution of the background signal (i.e. the instrumental and diffuse X-ray contribution unrelated to the source), which was obtained either through the use of archival observations or counts from the same observation but in a different energy band.

For the \textit{XMM-Newton} observations, we analyzed data from the EPIC pn camera \citep{PNcamera} and the two EPIC Metal-Oxide Semiconductor (MOS; \citealt{MOScamera}) cameras. These data were processed using the Science Analysis Software (SAS) 20.0.0 \citep{2004ASAS} and the latest calibration files. The EPIC events were cleaned with standard filtering expressions\footnote{For pn data: \texttt{$\#$XMMEA\_EP$\&\&$(FLAG==0)$\&\&$(PATTERN<=4)} and for MOS: \texttt{$\#$XMMEA\_EM$\&\&$(PATTERN<=12).}}. Point-like sources, identified through the SAS source detection task \texttt{emldetect}, were excluded by masking circular regions, each with a radius of 25$^{\prime\prime}$ (except for particularly bright sources where the radius was set to 40$^{\prime\prime}$).

For most \textit{XMM-Newton} observations, the pn background was extracted over a broad energy range of 0.5-15 keV. 
We excluded two intervals: the band containing significant ring emission (0.7-4 keV for GRB 221009A and GRB 160623A, and 0.8-2.2 keV for GRB 031203), and the 7.3-9.3 keV region. The latter contains fluorescence features from the detectors and their surrounding structure, whose intensity is not uniform across the pn camera\footnote{This band contains the emission lines corresponding to $Cu-K\alpha$, $Ni-K\alpha$, and $Zn-K\alpha$.}. For the MOS cameras, we adopted a similar strategy, extracting the background from the 4-12 keV range.

The background events extracted in this way are treated as an independent sample and analyzed following the same procedure adopted for the source regions. In particular, a pseudo-distance is assigned to the background events producing a background pseudo-distance distribution. Before subtraction, this distribution extracted from an energy range where no dust-scattering emission is observed must be rescaled by the ratio of the number of counts detected in the source and background energy ranges in a spatial region where no dust-scattering signal is expected at any energy.

The \textit{Chandra} data reduction was performed with the Chandra Interactive Analysis of Observations (\texttt{CIAO v4.15}, \citealt{2006SPIE.6270E..1VF}) software and the calibration data files \texttt{CALDB(v4.10.2)}. In order to reprocess the data, the \textsf{chandra$\_$repro} script was used, while the \textsf{blank-sky} script automatically selected the appropriate blank-sky background files in each observation. Point-like sources were identified using the \textsf{wavdetect} tool and excluded from the analysis.

\begin{table*}[ht]
	\centering
	\caption{Log of the \textit{Chandra} and \textit{XMM-Newton} observations of the GRBs.}
	\label{tableobsid}
	\begin{tabular}{llcccc} 
		\hline
		& Satellite/instrument &  Observation ID & Start time (UTC) & Stop time (UTC) & Exposure (ks)\\
		\hline
        GRB 221009A&\textit{XMM-Newton}/pn&  0913991601 (Obs2) & 2022-10-14T05:41:21 & 2022-10-14T22:52:21 &  $34.0$\\
        & \textit{Chandra}/ACIS &  27517 & 2022-10-17T16:53:32 & 2022-10-17T23:36:58 & 21.7\\
        &\textit{XMM-Newton}/pn& 0913991701 (Obs3) & 2022-10-30T04:56:37 & 2022-10-30T21:48:17 &  $34.0$\\
        &\textit{XMM-Newton}/pn& 0913991801 (Obs4)& 2022-11-01T07:21:37 & 2022-11-01T21:39:57 &  $35.0$\\
        &\textit{XMM-Newton}/pn& 0913991901 (Obs5)& 2022-11-10T16:09:14 & 2022-11-11T21:00:54 &  $87.8$\\
    	\hline

        GRB 160623A&\textit{XMM-Newton}/pn & 0781100201 & 2016-06-24T20:23:46 & 2016-06-25T12:53:464 &  $50.0$\\
        &\textit{XMM-Newton}/MOS&  &  &  &  $55.0$\\
     	\hline
        GRB 031203&\textit{XMM-Newton}/pn&  0158160201 & 2003-12-04T04:09:29 & 2003-12-04T20:19:40 &  $56.0$\\
        &\textit{XMM-Newton}/MOS&  &  &  &$56.0$\\
        &\textit{XMM-Newton}/pn& 0163360201 & 2003-12-06T18:11:07 & 2003-12-07T09:21:17 &  $37.0$\\
        &\textit{XMM-Newton}/MOS&  &  &  &$35.0$\\
        &\textit{Chandra}/ACIS &  05298 & 2004-01-22T21:17:36 & 2004-01-23T03:44:30 & 21.5\\
		\hline
     \end{tabular}
\end{table*}

\subsection{GRB 221009A}\label{sec2.1}
GRB 221009A was initially reported by the Burst Alert Telescope (BAT) aboard the \textit{Swift Neil Gehrels Observatory} on October 9, 2022, at 14:10:17 UT \citep{DichiaraGCN}. However, the GRB started about an hour before as observed for example by the \textit{Fermi} Gamma-ray Burst Monitor (GBM; \citealt{Lesage2023}) . This GRB is the brightest ever recorded and occurred near the direction of the Galactic plane, producing multiple X-ray dust scattering rings later observed by the X-ray Telescope (XRT) on Swift \citep{Williams2023, Vasilopoulos2023, Vaia2025}, as well as by \textit{XMM-Newton} \citep{Tiengo2023, Vaia2025}, the \textit{Imaging X-ray Polarimetry Explorer} (IXPE) \citep{Negro2023}, and \textit{Chandra}. For this study, we shall focus on data from \textit{XMM-Newton} and \textit{Chandra}, which are more sensitive to the relatively small and faint rings produced by distant dust.
\subsubsection{XMM-Newton observations}\label{sec2.1.1}
\textit{XMM-Newton} observed GRB 221009A on five occasions from October 11 to November 11 2022.
To minimize pile-up during the first observation, the pn and the two MOS cameras were operated in Timing mode, using the thick optical-blocking filter. In Timing mode, only the peripheral CCDs of the MOS cameras provide full imaging capabilities, covering a region from approximately $5^\prime$ to $15^\prime$ from the target position.
As a result, only the rings produced by clouds at distances smaller than $1.6\,\rm{kpc}$ (see equation \ref{eq:theta}) are visible in this observation \citep{Tiengo2023,Vaia2025}. This data set is not used in this work, as we focus on clouds at distances greater than $5\,\rm{kpc}$. The log of all the observations analysed in this work is reported in Table \ref{tableobsid}.

As a background for Obs2, we selected, for both cameras, data from the observation of the bright high-redshift quasar RBS 315 conducted in January 2013 (Observation ID: 0690900201). As in \citet{Vaia2025}, this observation was chosen due to the similarities of the central source with the GRB 221009A X-ray afterglow and because it occurred during a similar phase of the Solar Cycle, ensuring comparable contamination from the unfocused particle background \citep{GastaldelloSP}. We adopted this strategy for the background to subtract the contribution from the X-ray background, as well as from the GRB afterglow and its possible dust-scattering halo.
The approach outlined at the beginning of Section 2, was instead applied to the remaining three \textit{XMM-Newton} observations.
To maximize the signal-to-noise ratio of the rings, the analysis was confined to the 0.7--4 keV energy band, as in \citet{Tiengo2023} and \citet{Vaia2025}.

\subsubsection{Chandra observation}\label{sec2.1.2}
\begin{figure}[h!]
  \centering
  \includegraphics[width=0.5\textwidth]{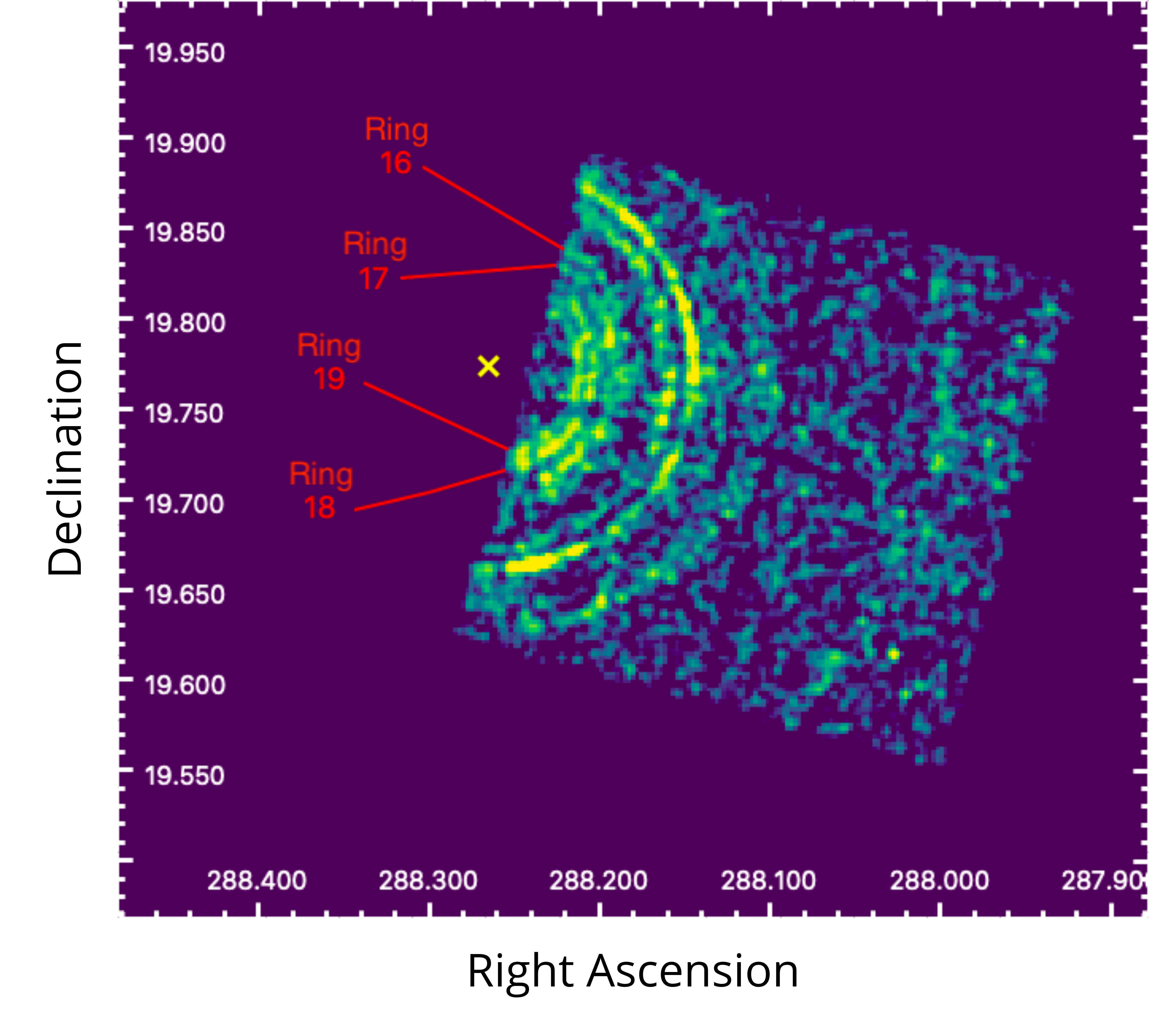}
  \caption{0.7--4 keV ACIS-I \textit{Chandra} image of GRB 221009A. The image is smoothed with a Gaussian kernel of $\sigma = 1''$ and the point sources are excluded. The yellow cross marks the position of the GRB afterglow. The red labels indicate the positions of the rings 16, 17, 18, and 19 produced by clouds at approximately 8, 10, 14 and 19 kpc, respectively.}
  \label{rings22109A}
\end{figure}

The \textit{Chandra} telescope started observing GRB 221009A on October 17, 2022 at 16:53:32 UT with an exposure time of 21.7 ks (Table \ref{tableobsid}). The data were acquired using the ACIS-I instrument in FAINT mode. As shown in Fig.\ref{rings22109A}, the rings are not fully covered and only a portion of them is visible since the observation was performed off-axis. This strategy was chosen to exclude the afterglow emission and allow the detection of the largest rings.

No significant background flares were detected and so the full exposure could be analysed.
As an alternative to blank-sky background files, we also selected events from the same observation, but in a hard energy band (4-10 keV), where no signal from X-ray dust-scattering is expected.
To maximize the signal-to-noise ratio of the scattered X-ray rings, the analysis was restricted to the 0.7--4 keV energy band as for the \textit{XMM-Newton} data.

\subsection{GRB 160623A}\label{sec2.2}
GRB 160623A was detected by the Fermi/GBM instrument on June 23, 2016, at 05:00:34.23 UT. This burst was also observed above 1 GeV by Fermi/LAT \citep{Vianello2016GCN} and detected by Konus-Wind \citep{Frederiks2016GCN}. Approximately 40 ks after the burst, Swift/XRT performed its first observation of the GRB afterglow, revealing a dust-scattering ring with a radius of about $3.5^\prime$ \citep{Tiengo2016GCN}, attributed to a dust cloud located at approximately 800 pc (see Eq. \ref{eq_distance}). In a subsequent \textit{XMM-Newton} observation six X-ray dust-scattering rings were identified \citep{Pintore2017}.
The \textit{XMM-Newton} satellite started observing GRB 160623A approximately two days later on June 24, 2016 at  20:23:46 UT, employing all EPIC cameras in full-frame mode for a duration of 16 hours (Table \ref{tableobsid}). No more \textit{XMM-Newton} or \textit{Chandra}, observations of this field are available.

\subsection{GRB 031203}\label{sec2.3}
On December 3, 2003, at 22:01:28 UT, GRB 031203 was identified by the Imager aboard the INTEGRAL satellite (IBIS; \citealt{Gotz2003GCN}). This GRB had a single peak with a duration of 30 s and a peak flux of $1.3\,\times\,10^{-7}\rm{erg\,s^{-1}\,cm^{-2}}$ in the $20\,-\,200$ keV energy range. At the time of the burst, the \textit{Swift} satellite was not yet operational. The \textit{XMM-Newton} observatory reacted immediately, while \textit{Chandra} performed its first observations more than a month after the burst.

\subsubsection{XMM-Newton observations}\label{sec2.3.1}

A 58 ks observation of the GRB 031203 field with \textit{XMM-Newton} started about six hours after the burst, on December 4, 2003, at 04:09:29 UT (Table \ref{tableobsid}). The \textit{XMM-Newton} satellite captured the GRB on-axis, utilizing all EPIC cameras in full-frame mode. 

On December 6, 2003, GRB 031203 was serendipitously observed at an off-axis angle of $\sim 10^{\prime}$ during a Zeta Puppis pointing (Table \ref{tableobsid}). Following Eq. \ref{eq:theta}, the radius of the ring detected by \citet{Feng2010}, which was generated by a cloud at a distance of 9.9 kpc, is expected to be $\sim 2.4^{\prime}$ after three days, remaining fully within the Field of View of all three instruments. To maximize the signal-to-noise ratio of the rings, the analysis was confined to the 0.8--2.2 keV energy band, as in \citet{Tienghetti2006}.

\subsubsection{Chandra observation}\label{sec2.3.2}
The \textit{Chandra} telescope started observing GRB 031203 on January 22, 2004 at 21:17:36 UT with an exposure time of 21.5 ks (Table \ref{tableobsid}). A 30 ks \textit{Chandra} observation taken on April 18, 2004 was not considered because no detectable X-ray rings were expected more than 100 days after the GRB.

The data were acquired using both ACIS-I and ACIS-S chips in VFAINT mode. No significant background flares were detected during the observation. As in the \textit{XMM-Newton} analysis, we focused on the 0.8--2.2 keV energy. The background blank-sky subtraction was performed separately for CCD S4 and for CCDs I2, I3, S2 and S3, due to the different exposure time of CCD S4 blank-sky data. To provide additional background estimates, the original event file was filtered both in the 2.2--5 keV and 4--10 keV high energy bands, where negligible dust-scattered X-ray emission is expected.

\section{Data analysis and results}\label{sec3}
In this study, we analyze the X-ray dust-scattering rings of GRB 221009A, GRB 160623A, and GRB 031203, with a particular focus on determining the distances to dust clouds located beyond 5 kpc. For these GRBs, we applied the pseudo-distance distribution method \citep{Tienghetti2006} to measure accurately the distances of previously discovered dust structures and to search for additional X-ray rings produced by dust in the Perseus, Outer, and OSC arms.

To produce a pseudo-distance distribution, for each detected event, we computed the quantities:
\begin{equation}
    t_{i} = T_{i} - T_0,
\end{equation}

\begin{equation}
     \theta_{i}^2 = (x_{i} - X_{\rm{GRB}})^2 + (y_{i} - Y_{\rm{GRB}})^2,
\end{equation}
where $x_{i}$, $y_{i}$, and $T_{i}$ are the detector coordinates and the time of arrival of the i-th event, and $T_0$, $X_{\rm{GRB}}$, and $Y_{\rm{GRB}}$ are the GRB time and detector coordinates, respectively.
Using these new coordinates, we computed the pseudo-distance defined as:
\begin{equation}
\label{eq_distance}
D_{i} = 2ct_{i} / \theta_{i}^2 = 827\,t_{i}\, [\rm{s}] \,\theta_i^{-2}\,[\rm{arcsec}]\, \rm{pc} 
\end{equation}
For background events, this quantity does not represent a true distance. However, photons from an expanding ring have $D_{i}$ values that cluster around the distance of the dust-scattering cloud, resulting in a ring appearing as a peak centered at the distance of the dust layer in the pseudo-distance distribution.

The data analysis for GRB 221009A, GRB 160623A, and GRB 031203 is detailed in the following subsections. 
\subsection{GRB 221009A}
For GRB 221009A, we created histograms of the pseudo-distances for all the observations and their respective backgrounds, but we limited our analysis only to pseudo-distances larger than 6 kpc. To indicate the peaks and the corresponding rings in the X-ray images, we adopted the nomenclature introduced in \citet{Tiengo2023}, where Ring 16, 17, 18, and 19 were identified in this distance range.

\begin{figure*}[h]
\centering
\begin{subfigure}{0.4\textwidth}
\includegraphics[width=\linewidth]{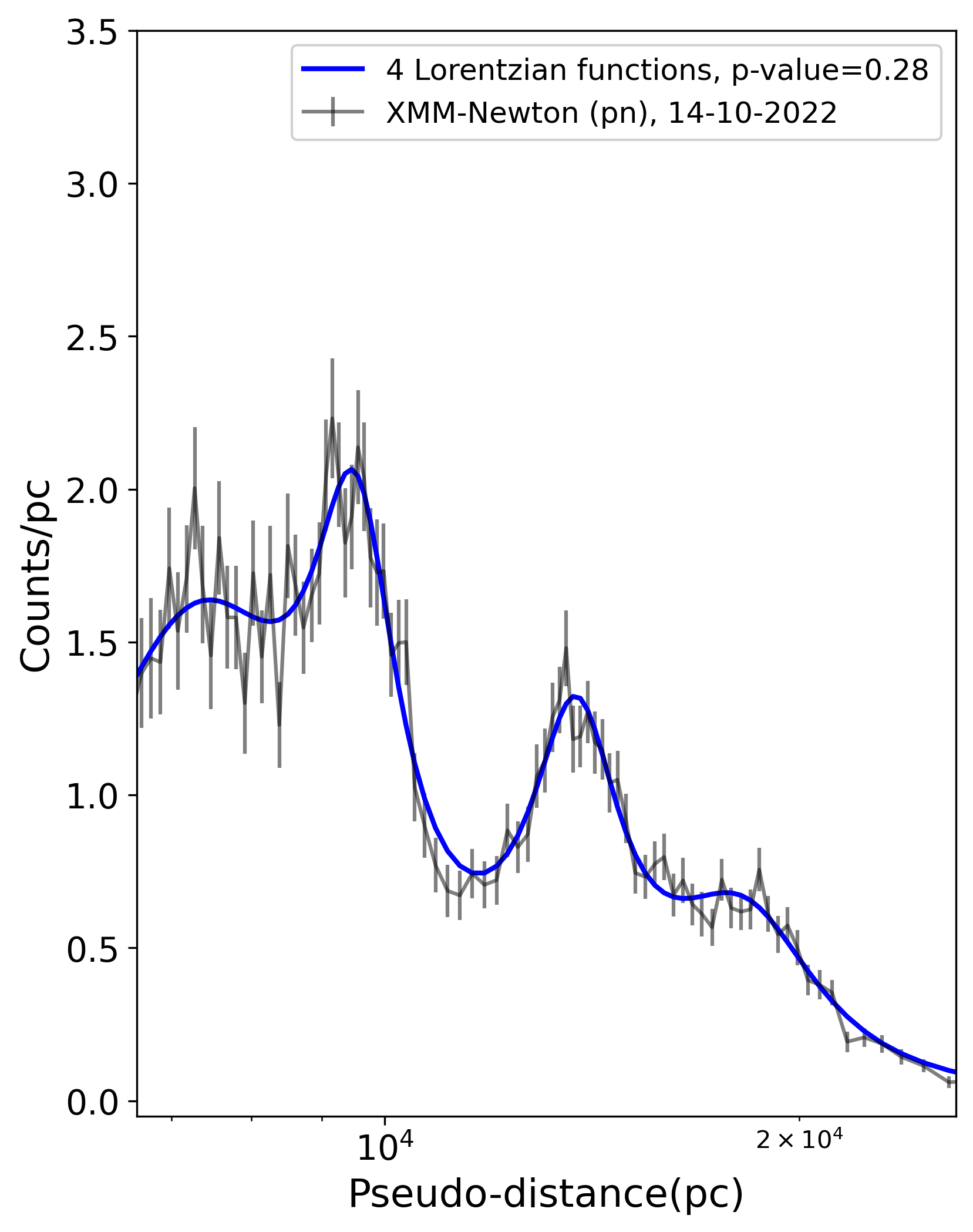}
\end{subfigure}
\begin{subfigure}{0.3775\textwidth}
\includegraphics[width=\linewidth]{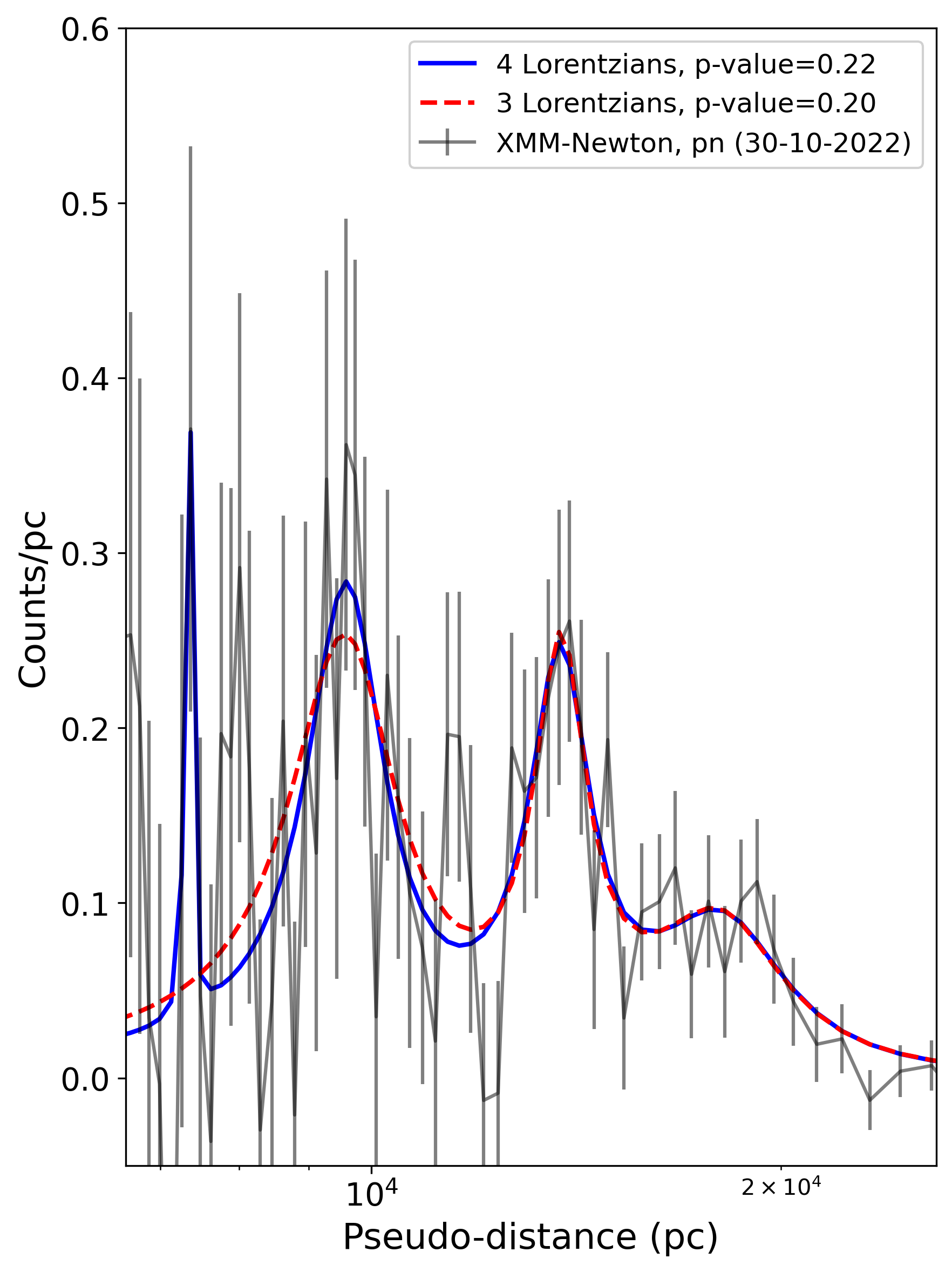}
\end{subfigure}
\begin{subfigure}{0.4\textwidth}
\includegraphics[width=\linewidth]{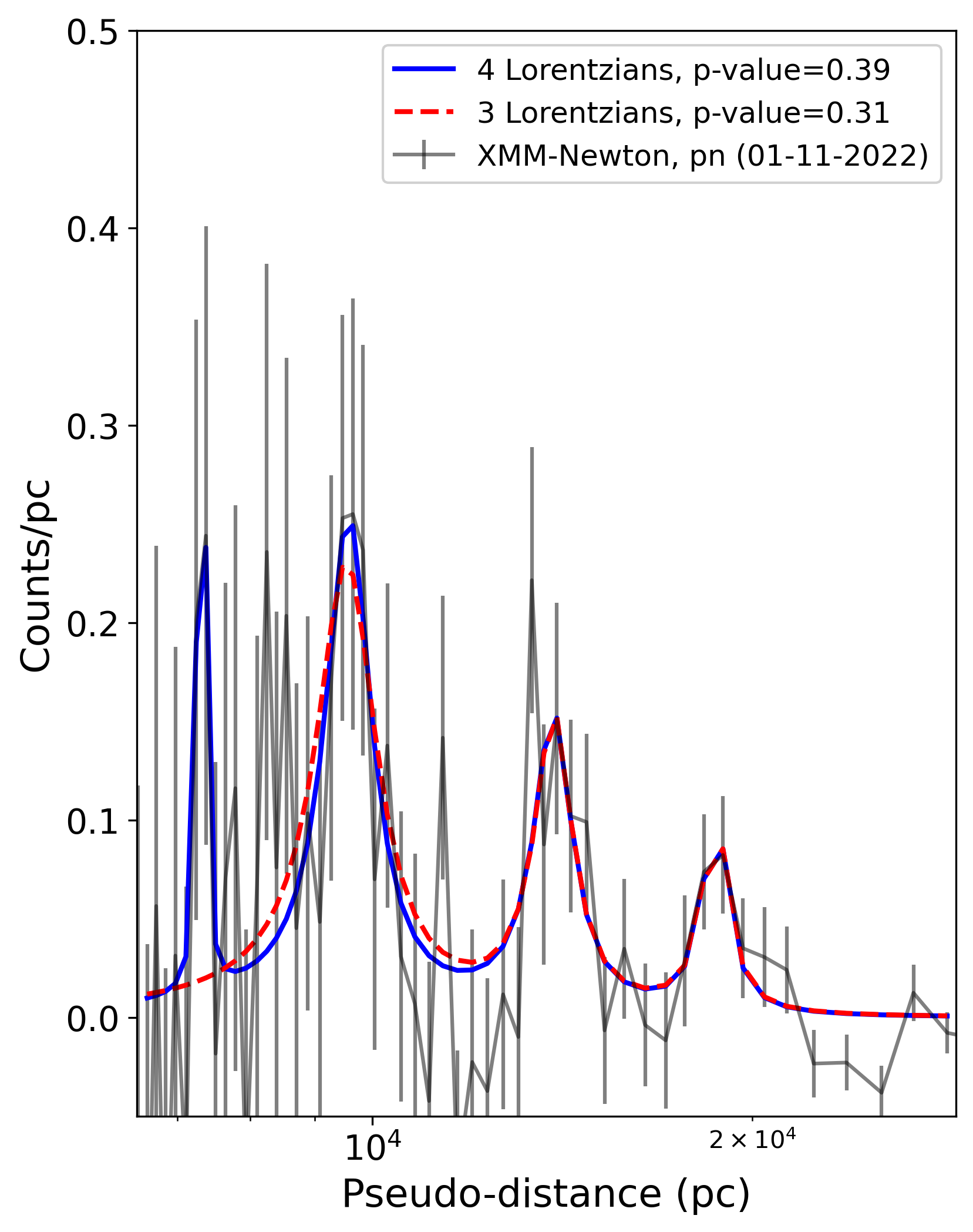}
\end{subfigure}
\begin{subfigure}{0.4\textwidth}
\includegraphics[width=\linewidth]{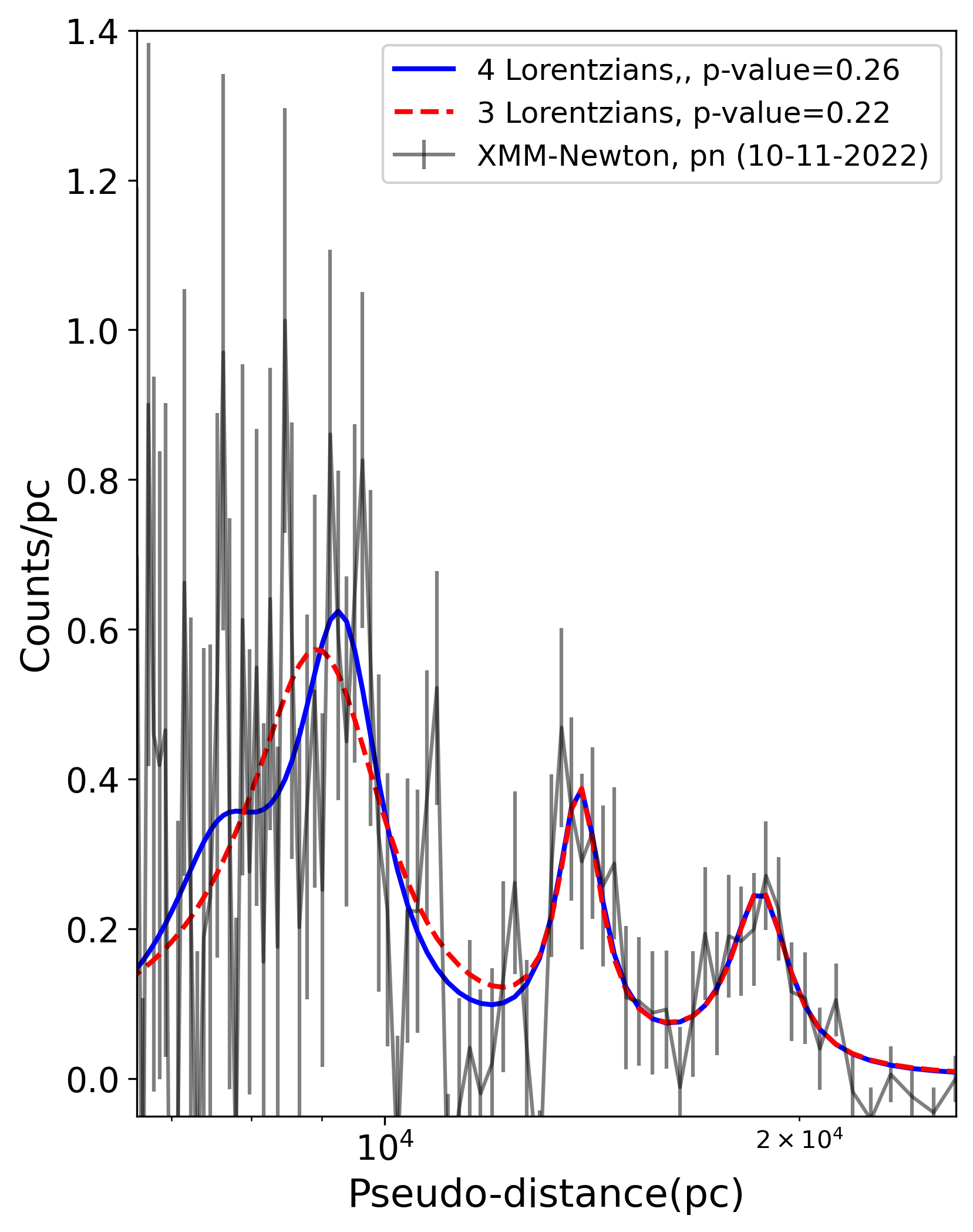}
\end{subfigure}
\caption{Pseudo-distance distributions in the 0.7--4 keV energy band for the GRB 221009A observations performed by \textit{XMM-Newton} on October 14 (Obs2, top-left panel), October 30 (Obs3, top-right panel), November 1 (Obs4, bottom-left panel), and November 10 (Obs5, bottom-right panel). The red dashed and blue solid lines indicate the best-fit models with 3 and 4 Lorentzian functions, respectively.}
\label{221009xmm}

\end{figure*}

For \textit{XMM-Newton} observations, we conducted separate analyses for the data collected by the MOS and pn cameras.
After verifying that the results were consistent and that the pn camera provided smaller uncertainties, we chose to present only the pn data.
For Obs2, the background-subtracted pseudo-distance was fit in the range from 6 to 30 kpc using a model comprising the sum of four Lorentzian functions , resulting in a good fit (p-value\footnote{The p-value is computed from the $\chi^2$ distribution under the null hypothesis that the model correctly describes the data. Throughout this work, we adopt a p-value threshold of 0.01, below which models are rejected as statistically inconsistent with the null hypothesis.} = 0.28, see the top-left panel of Fig. \ref{221009xmm}). Each Lorentzian component corresponds to a distinct dust cloud, identified in \citet{Tiengo2023} for this GRB within the distance range fitted. The distances of these peaks correspond to those of the rings from 16 to 19 reported in \citet{Tiengo2023}.
Similarly, for Obs3 and Obs4, the pn background-subtracted pseudo-distances were adequately fit (p-value = 0.20 for Obs3 and p-value = 0.31 for Obs4) by the sum of three Lorentzian functions, corresponding to Ring 17, 18 and 19 (see Fig. \ref{221009xmm}).

To check if Ring 16 could be detected in these observations as well, we attempted a fit over the same distance range using the sum of four Lorentzian functions. 
Although the model provided good fits in both cases (Fig. \ref{221009xmm}), adding an extra Lorentzian did not improve the results, and the best-fit parameters of the three peaks remained consistent with those obtained from the former.

For Obs5, the data were well fitted with three Lorentzians (Rings 17-19, p-value = 0.22). A four-Lorentzian model was also tested to include Ring 16, but the extra peak's distance and width were unconstrained and thus fixed to Obs2 values. This addition did not improve the fit (p-value = 0.26) and left the other parameters unchanged (Fig. \ref{221009xmm}, bottom-right).

For the \textit{Chandra} observation, we derived the histogram of the pseudo-distances and tested both methods for background subtraction. 

The blank-sky background was simply rescaled by the ratio of the source and background exposure times, while the same observation at high energies was normalized by the ratio of the counts extracted in the 0.7--4 keV and 4--10 keV energy bands from a region free of scattering-ring contamination, corresponding to the 1.3-1.7 kpc distance interval. The background-subtracted data were fit with a model consisting of the sum of four Lorentzian functions over the 6.2-25 kpc distance range. The best-fit results are reported in Tab.\ref{confrontobkg}.

\begin{figure}[htbp]
  \centering
  \includegraphics[width=0.4\textwidth]{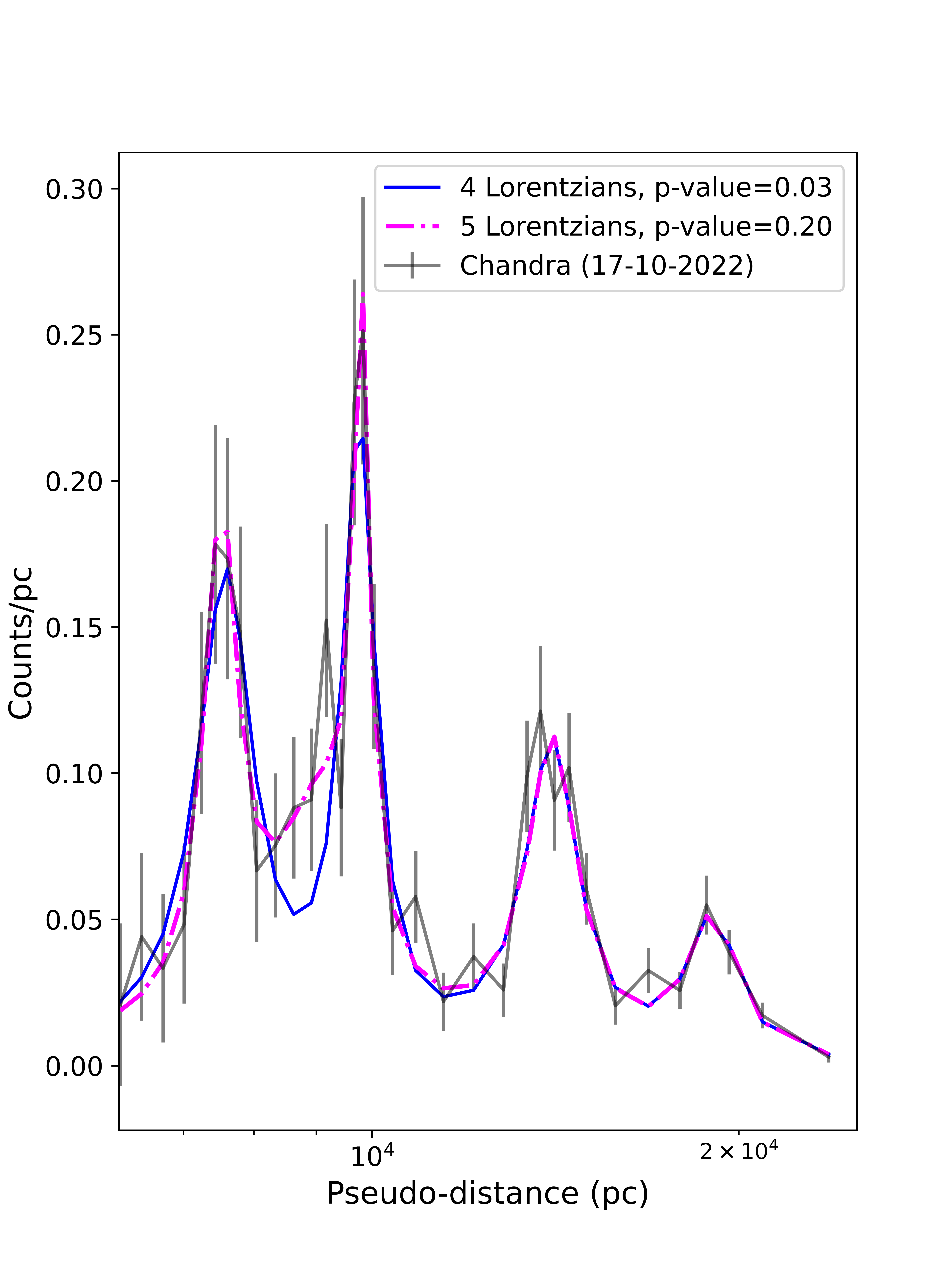}
  \caption{Pseudo-distance distribution in the 0.7--4 keV energy band for the GRB 221009A \textit{Chandra} observation. The background was subtracted from blank-sky observations.
  The blue line indicates the best-fit model with four Lorentzian functions, while the magenta line represents the best-fit model with five Lorentzian functions.}
  \label{221009blank}
\end{figure}

Since the fit obtained with the blank-sky background subtraction was only marginally acceptable, we also added a fifth Lorentzian function to the model. As can be seen in Fig.\ref{221009blank} and Tab.\ref{confrontobkg}, the additional Lorentzian accounts for a small excess between the first two peaks, with no significant effects on the positions of the four main peaks.

\renewcommand{\arraystretch}{1.6}
\setlength{\tabcolsep}{4pt}
\begin{table}
    \centering
    \caption{Best-fit parameters of the pseudo-distance distribution in the 0.7--4 keV energy band for the GRB 221009A \textit{Chandra} observation.}
    \begin{tabular}{lcccc}
        \noalign{\hrule height 0.8mm}
        & Ring 16 & Ring 17 & Ring 18 & Ring 19 \\
        \noalign{\hrule height 0.8mm}
        \multicolumn{5}{l}{\textbf{Blank-Sky (4 Lorentzians)}} \\
        D [kpc] & $7.58 \pm 0.07$ & $9.76 \pm 0.05$ & $14.0 \pm 0.1$ & $19.0 \pm 0.2$ \\
        \hline
        Width [kpc]    & $0.5 \pm 0.1$ & $0.36 \pm 0.07$  & $0.9 \pm 0.1$  & $1.1 \pm 0.3$ \\
        \hline
        Counts        & 124          & 119          & 145           & 86 \\
        \hline
        $\chi^2$/d.o.f. & \multicolumn{4}{c}{$36.62/22$} \\
        \hline
        p-value       & \multicolumn{4}{c}{0.026} \\
        \noalign{\hrule height 0.8mm}
        \multicolumn{5}{l}{\textbf{4--10 keV background (4 Lorentzians)}} \\
        D [kpc] & $7.66 \pm 0.08$ & $9.75 \pm 0.05$ & $13.9 \pm 0.1$ & $19.0 \pm 0.2$ \\
        \hline
        Width [kpc]    & $0.4 \pm 0.1$ & $0.33 \pm 0.07$  & $0.8 \pm 0.1$  & $1.1 \pm 0.3$ \\
        \hline
        Counts        & 93           & 106          & 129           & 76 \\
        \hline
        $\chi^2$/d.o.f. & \multicolumn{4}{c}{$28.65/22$} \\
        \hline
        p-value       & \multicolumn{4}{c}{0.155} \\
        \noalign{\hrule height 0.8mm}
        \multicolumn{5}{l}{\textbf{Blank-Sky (5 Lorentzians)}} \\
        D [kpc] & $7.53 \pm 0.06$ & \makecell{$9.1 \pm 0.4$ \\ $9.80 \pm 0.04$} & $14.1 \pm 0.1$ & $19.0 \pm 0.2$ \\
        \hline
        Width [kpc]    & $0.3 \pm 0.1$ & \makecell{$1.0 \pm 0.4$ \\ $0.16 \pm 0.07$}  & $0.8 \pm 0.1$  & $1.1 \pm 0.3$ \\
        \hline
        Counts        & 77           & \makecell{129 \\ 56}          & 139           & 85 \\
        \hline
        $\chi^2$/d.o.f & \multicolumn{4}{c}{$23.89/19$} \\
        \hline
        p-value       & \multicolumn{4}{c}{0.200} \\
        \noalign{\hrule height 0.8mm}
    \end{tabular}
    \tablefoot{Two background-subtraction methods (blank-sky or
4--10 keV background) and two models (4 or 5 Lorentzians) were
adopted.}
    \label{confrontobkg}
\end{table}

\begin{figure*}[h]
\centering
\begin{subfigure}{0.48\textwidth}
    \centering
    \includegraphics[width=\textwidth]{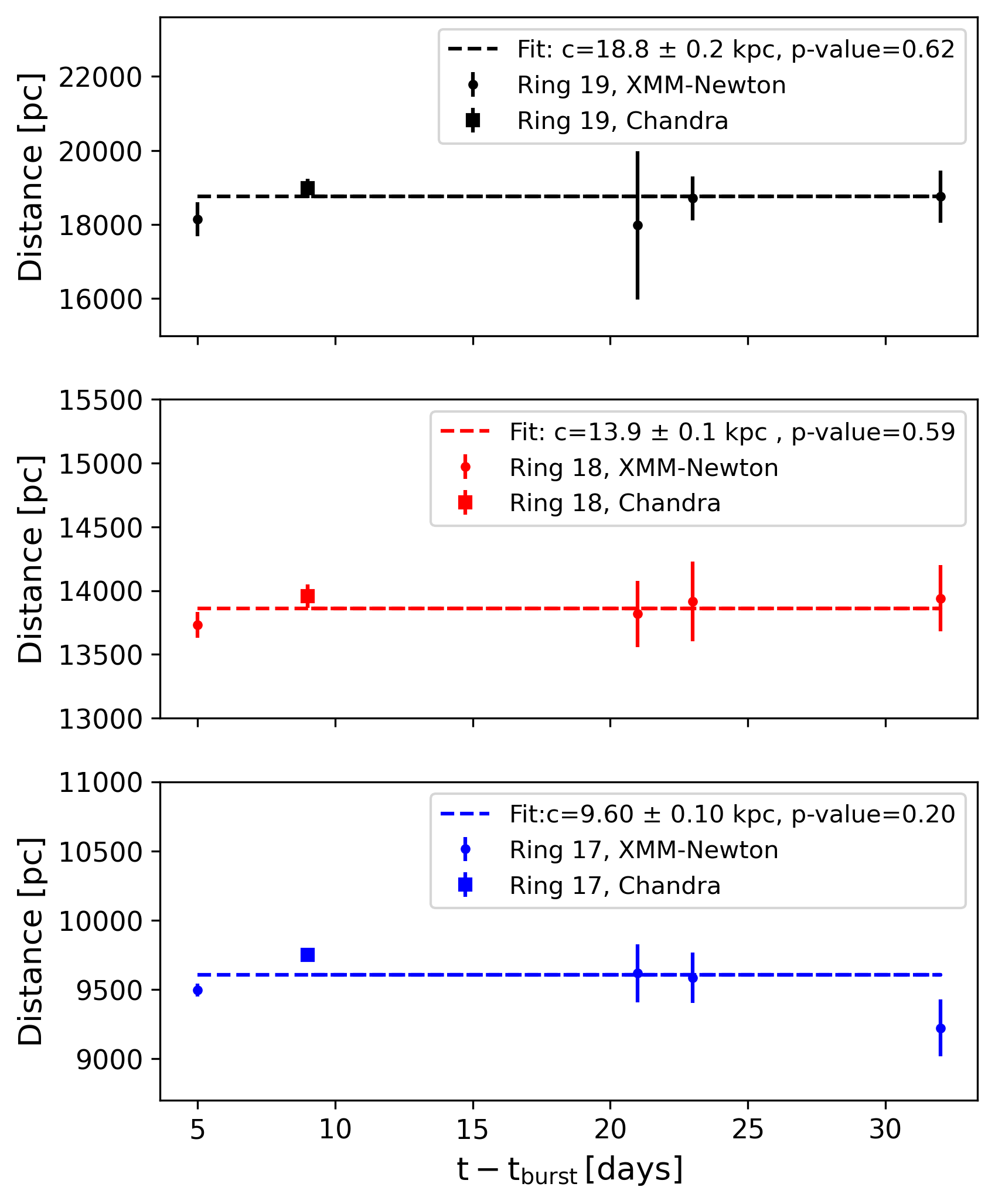}
    \label{fig:distances}
\end{subfigure}
\hfill
\begin{subfigure}{0.48\textwidth}
    \centering
    \includegraphics[width=\textwidth]{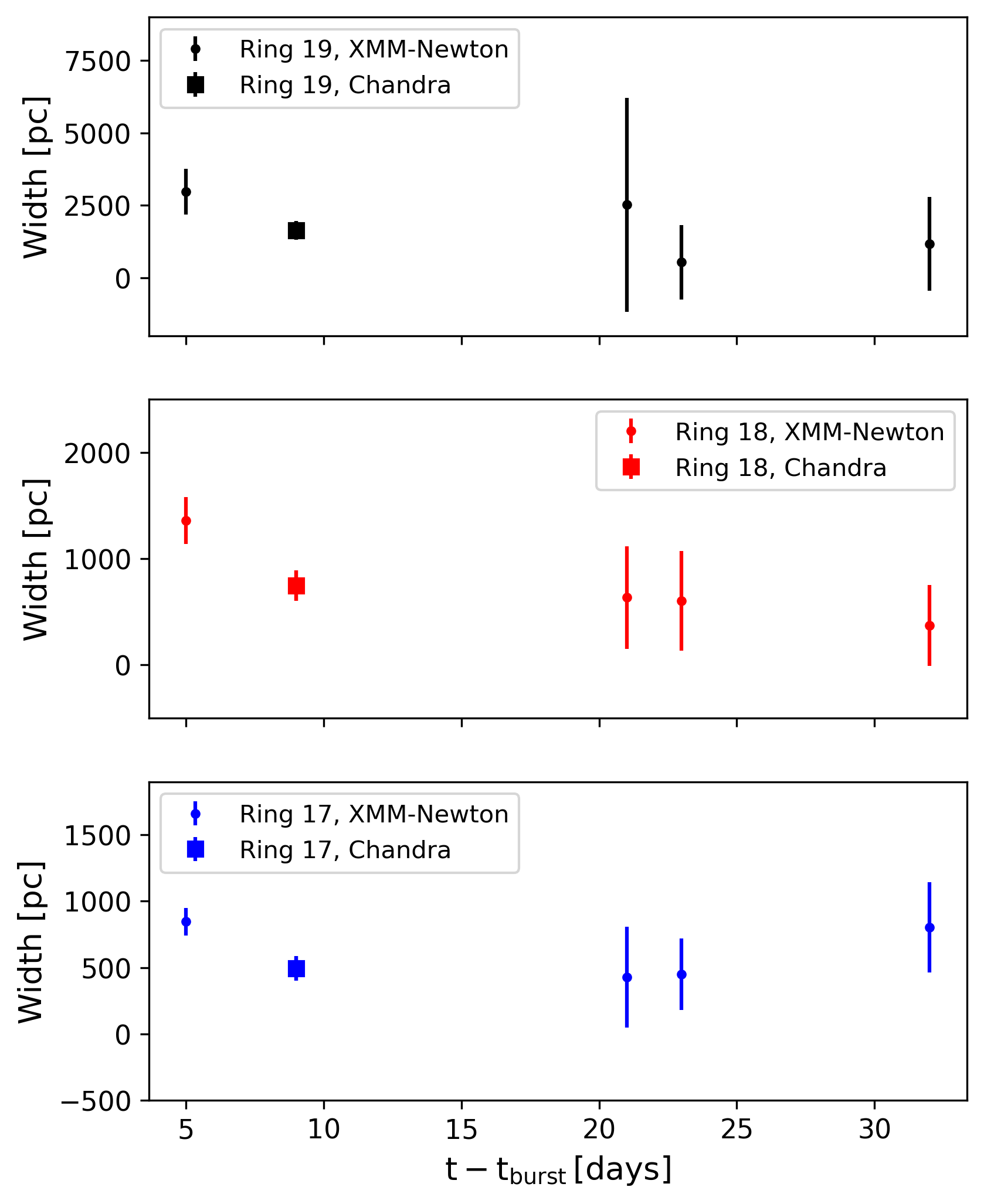}
    \label{fig:widths}
\end{subfigure}
\caption{Distances (left panels) and widths (right panels) of Ring 19 (upper panels), Ring 18 (central panels), and Ring 17 (bottom panels) determined from the \textit{XMM-Newton} (circles) and \textit{Chandra} (squares) observations of GRB 221009A. The dashed line represents the best-fit with a constant.}
\label{fig:results}
\end{figure*}

The distances and widths of the dust structures producing Ring 17, 18 and 19 derived from the \textit{XMM-Newton} and \textit{Chandra} observations are shown in Fig. \ref{fig:results}. 
 
A fit with a constant to the distances obtained from the different observations reproduces the data well, indicating that all the measurements -- despite being taken with different instruments and at different epochs -- are consistent. Similarly, the widths derived from the three late-time \textit{XMM-Newton} observations are in agreement with each other and with the value found by \textit{Chandra}. The width measured in Obs2 is significantly larger, but this can be explained by the fact that in this earlier observation the rings have smaller radii and the peak width is strongly affected by the instrumental point-spread function. The consistency among the other four observations demonstrates that the measured width corresponds to the intrinsic size of the dust cloud and is not dominated by the instrumental spatial resolution.

\subsection{GRB 160623A}

The X-ray dust scattering rings around GRB 160623A were previously studied by \citet{Pintore2017}, who identified six dust scattering rings, including one at a distance greater than 5 kpc, specifically at $5079\pm64$ pc. In \citet{Pintore2017}, the pseudo-distance method was also used to analyze the rings; however, in that case, the background was phenomenologically fit with the sum of two power-law functions, whereas in our analysis, we subtracted the background before fitting.
For this GRB, after having verified the consistency of data from all three cameras\footnote{We also included data from the MOS1 camera despite the damage sustained by two of its external CCDs early in the mission, as we are only interested in features at distances greater than 5 kpc, corresponding to angles smaller than  $\simeq 3^\prime$, which are fully covered by the central chip.}, we combined all data to minimize uncertainties.
We then derived the pseudo-distance distribution from all cameras for both the observation and the background (obtained as described in Section \ref{sec2}). The two distributions were subtracted, with normalization applied in a region free from X-ray dust scattering contamination, following the same approach as for GRB 221009A (in this case, the distance interval was 2--3 kpc).
From the background-subtracted pseudo-distance distribution at distances greater than 5 kpc, we confirm the peak previously identified by \citet{Pintore2017} at $\rm{D_1= 5125 \pm 29\,\rm{pc}}$ ($\rm{\sigma_1}=410 \pm 47$ pc). Additionally, we detect two more peaks at distances $\rm{D_2}=6909 \pm 56$ pc and $\rm{D_3}= 9908 \pm 624$ pc, with widths of $\rm{\sigma_2}=299 \pm 91$ pc and $\rm{\sigma_3}= 1027 \pm 190$ pc, respectively.

\begin{figure}[h]
\centering
\includegraphics[width=0.4\textwidth]{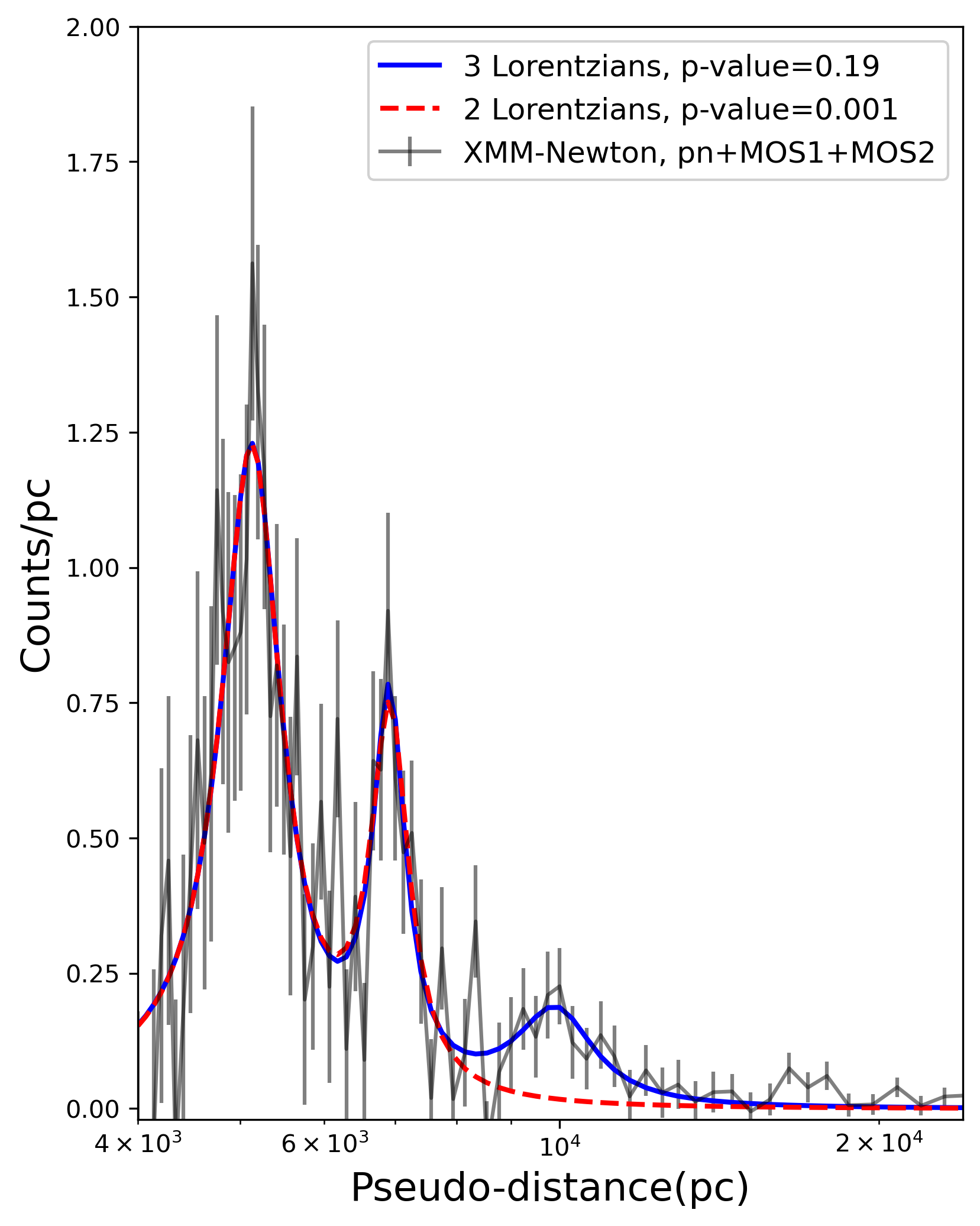}
\caption{Pseudodistance distributions in the 0.7--4 keV energy band for the GRB 160623A \textit{XMM-Newton} observations of June 24 between 4 and 20 kpc. The blue solid line indicates the best-fit model with 3 Lorentzian functions. The red dashed line indicates the best-fit model with 2 Lorentzian functions.}
\label{160623}
\end{figure}

When modeling the data with the sum of only two Lorentzian functions, centered at $\rm{D_1}$ and $\rm{D_2}$, we obtain a p-value of 0.001, whereas a model incorporating three Lorentzian functions yields a p-value of 0.19 (see Figure \ref{160623}).

\subsection{GRB 031203}
The X-ray dust scattering rings around GRB 031203 have been extensively studied over the years. \citet{Vaughan2004} identified two dust scattering rings at distances of less than 2 kpc, a finding later confirmed by \citet{Tienghetti2006} using the pseudo-distance method.
A subsequent analysis of the same data by \citet{Feng2010} reported the discovery of a third ring at a distance $9.94\,\pm\,0.39\,\rm{kpc}$.

\begin{figure}
\centering
\begin{subfigure}{0.4\textwidth}
    \centering
    \includegraphics[width=\textwidth]{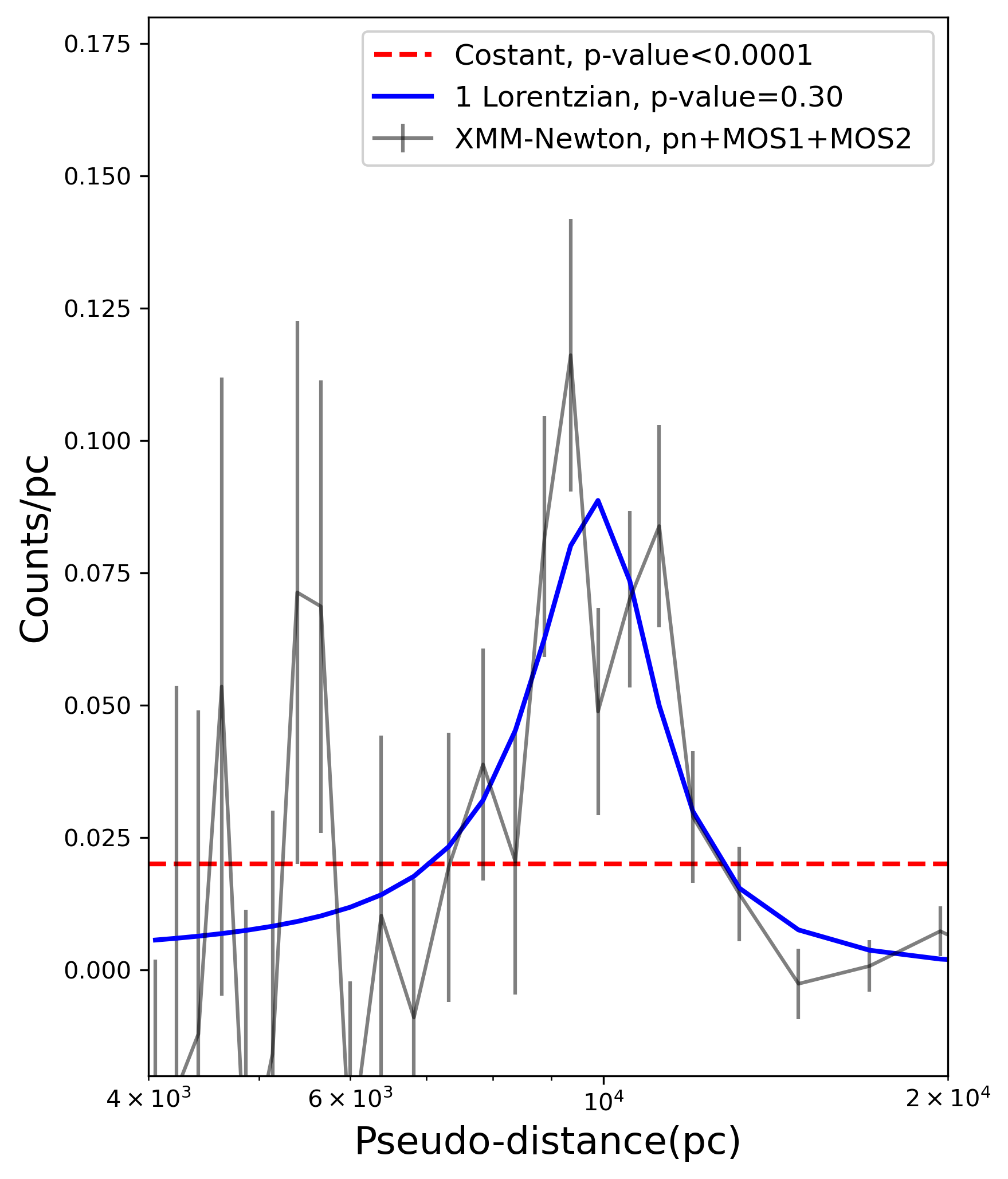}   
\end{subfigure}
\hfill
\begin{subfigure}{0.4\textwidth}
    \centering
    \includegraphics[width=\textwidth]{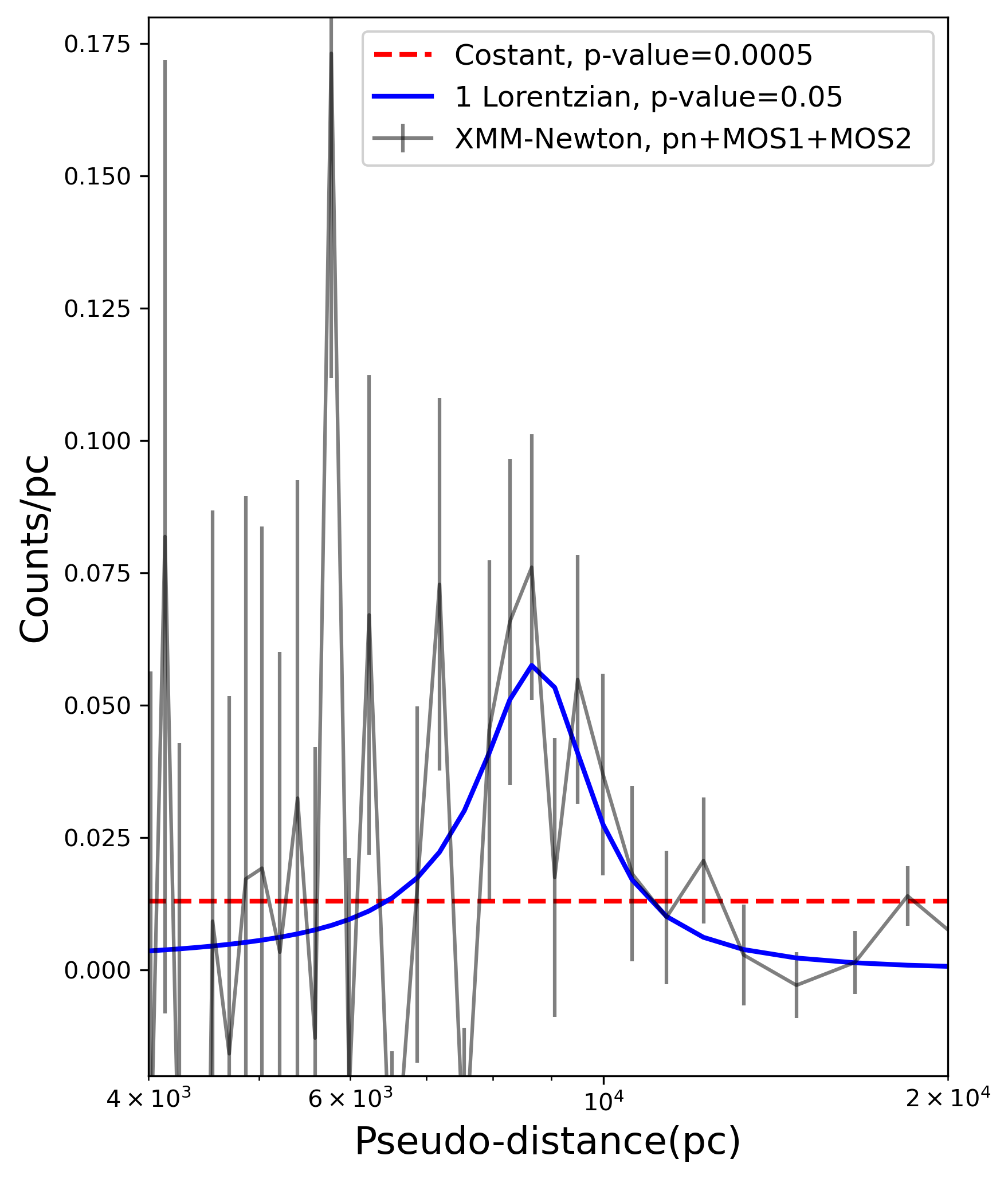}
\end{subfigure}
\caption{Pseudodistance distributions in the 0.8--2.2 keV band for the \textit{XMM-Newton} observations of GRB~031203. The blue solid line shows the best-fit with a Lorentzian function, and the red dashed line the best-fit with a constant.
Upper panel: December~4. Bottom panel: December~6.}
\label{fig:grb031203}
\end{figure}

In this work, we reanalyze the first \textit{XMM-Newton} observation and, for the first time, apply the pseudo-distance distribution method to both the \textit{Chandra} data and the second \textit{XMM-Newton} observation of this GRB. For \textit{XMM-Newton}, after confirming consistency across all three cameras -- following a procedure similar to that used for GRB 160623A -- we combined the data of each observation to reduce statistical uncertainties. We then derived the pseudo-distance distributions for both observations and subtracted their corresponding backgrounds (as described in Section \ref{sec2}).
The background distributions were normalized in a region free from X-ray dust scattering contamination, specifically between 4 and 7 kpc, in line with the method used for GRB 221009A and GRB 160623A.
By fitting the background-subtracted pseudo-distance distribution of the first observation in the 4--25 kpc range, we confirmed the presence of the peak identified, with a different method, by \citet{Feng2010}. Our analysis yielded a dust cloud distance of $9.9\,\pm\,0.4\,\rm{kpc}$ and a width of $\sigma = 1.5\,\pm\,0.6\,\rm{kpc}$, with a fit p-value of 0.30 (see upper panel Figure \ref{fig:grb031203}). 
The same peak is also visible in the second \textit{XMM-Newton} observation, though with lower significance. Fitting its background-subtracted pseudo-distance distribution, we found a peak at $8.7\,\pm\,0.8\,\rm{kpc}$ with a width of $\sigma = 1.2\,\pm\,0.7\,\rm{kpc}$. This fit returned a p-value of 0.05, while a fit using a constant model resulted in a much lower p-value of 0.0005 (see the bottom panel of Figure \ref{fig:grb031203}). The weighted average of these two measurements provides a distance of $9.7\,\pm\,0.4\,\rm{kpc}$.

For the \textit{Chandra} observation, we derived histograms of the pseudo-distances both subtracting blank-sky and high-energy backgrounds, but no significant dust-scattered emission was detected. 
In the latter case, the background pseudo-distance values were normalized by the ratio of the total counts in the 0.8--2.2 keV band to those in the 2.2--5 keV range, extracted from a region free of scattering ring contamination (corresponding to distances in the 12--15 kpc interval). The fit of the background-subtracted distribution with a constant was perfectly consistent with no signal in the 3.5--20 kpc distance range ($y=-0.007\pm 0.004$, $\chi^2/dof = 27.43/35$, p-value $=0.82$).

\section{Discussion}

By analyzing the expanding X-ray scattering rings of GRB 221009A, GRB 160623A, and GRB 031203, we have been able to map the outer structure of the Milky Way with an unprecedented level of precision. Unlike the most commonly used methods that rely on kinematic models, the dust scattering rings provide a direct geometric measurement of the distance to dust clouds along the line of sight. Owing to the good angular resolution of our X-ray data, we can measure not only the distance but also the spatial thickness of the spiral arms, as the shape of the peaks in the pseudo-distance distributions reflects the radial distribution of dust within each arm. 

In Fig. \ref{Galaxy}, we overlaid the 3 directions sampled by the selected GRBs on the Milky Way picture reported in \citet{Sun2024}, which is based on the \citet{Reid2019} rotation curve. 

As indicated by the cyan dots, our analysis allows us to probe the structure of the Perseus, Outer, and OSC arms.

In the following, we compare our results with the distances of Galactic structures derived from \ion{H}{I}/CO data in the same directions, assuming different Galactic rotation curves, and we discuss their implications on the characterization of each spiral arm.

\subsection{Comparison with \ion{H}{I}/CO Data and Kinematic Models}
\begin{figure}[h!]
    \centering
    \begin{subfigure}[b]{0.42\textwidth} 
        \centering
        \includegraphics[width=\textwidth]{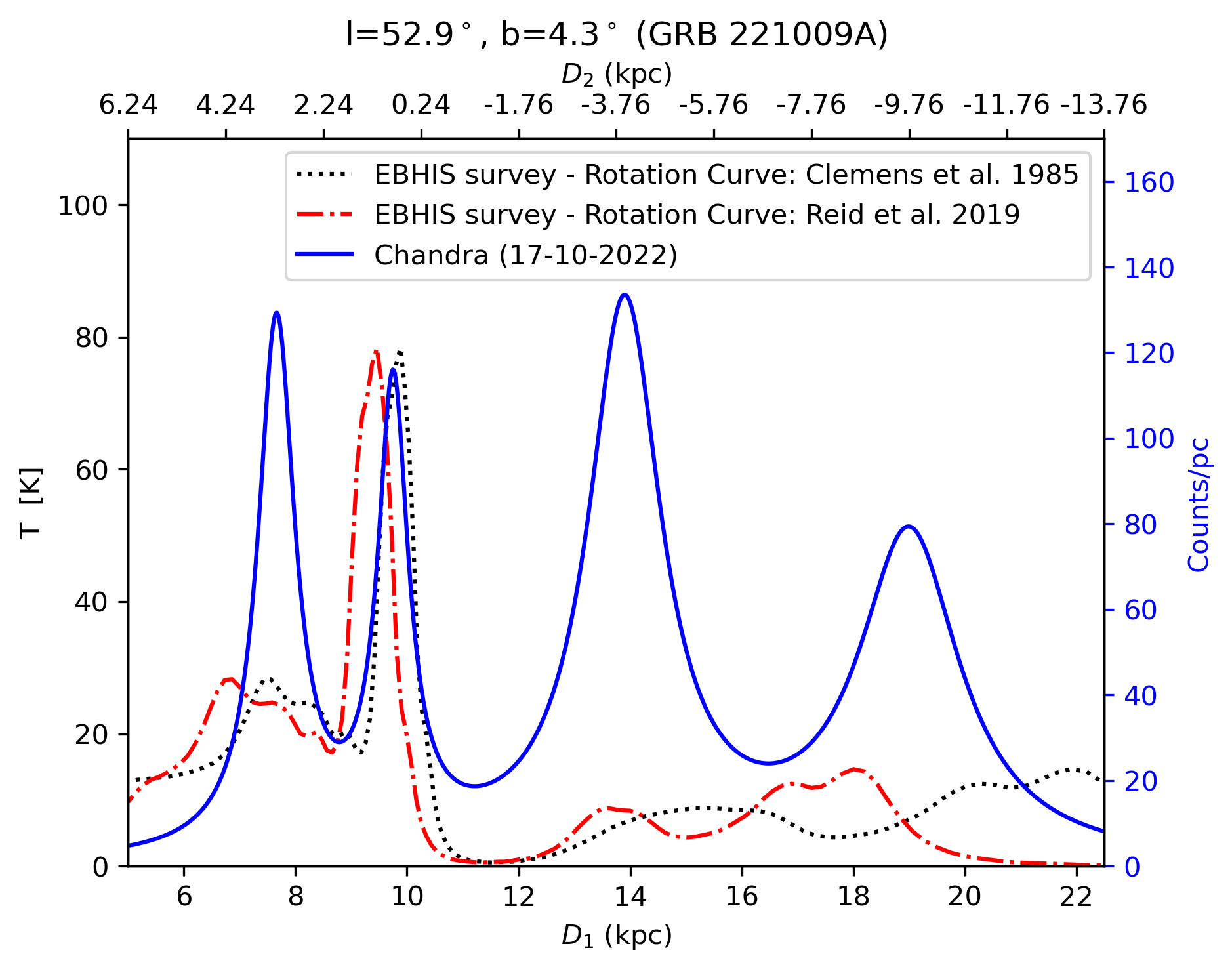}
        \label{EBHIS1}
    \end{subfigure}
    \hfill
    \begin{subfigure}[b]{0.42\textwidth}
        \centering
        \includegraphics[width=\textwidth]{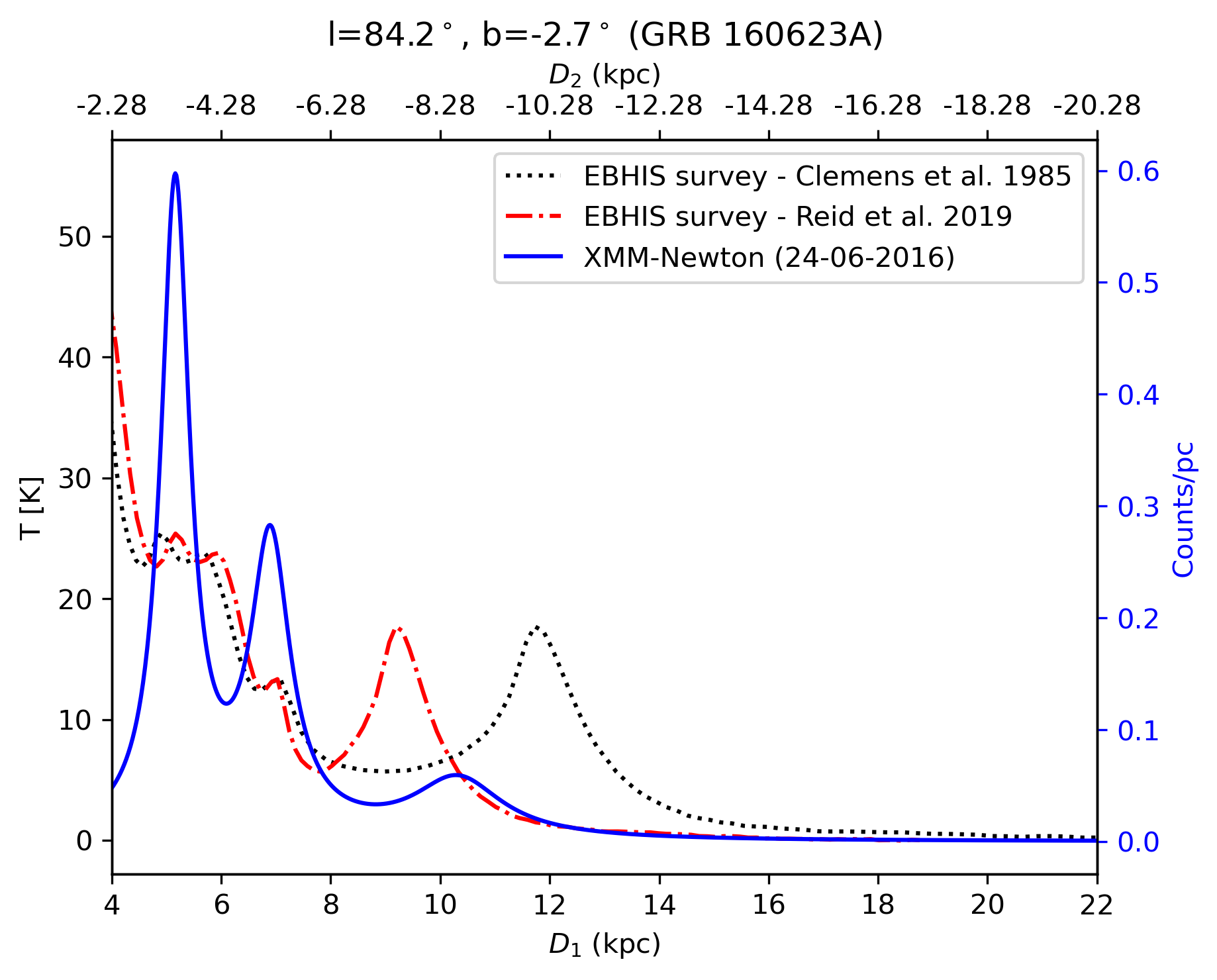}
        \label{EBHIS2}
    \end{subfigure}
    \hfill
    \begin{subfigure}[b]{0.42\textwidth} 
        \centering
        \includegraphics[width=\textwidth]{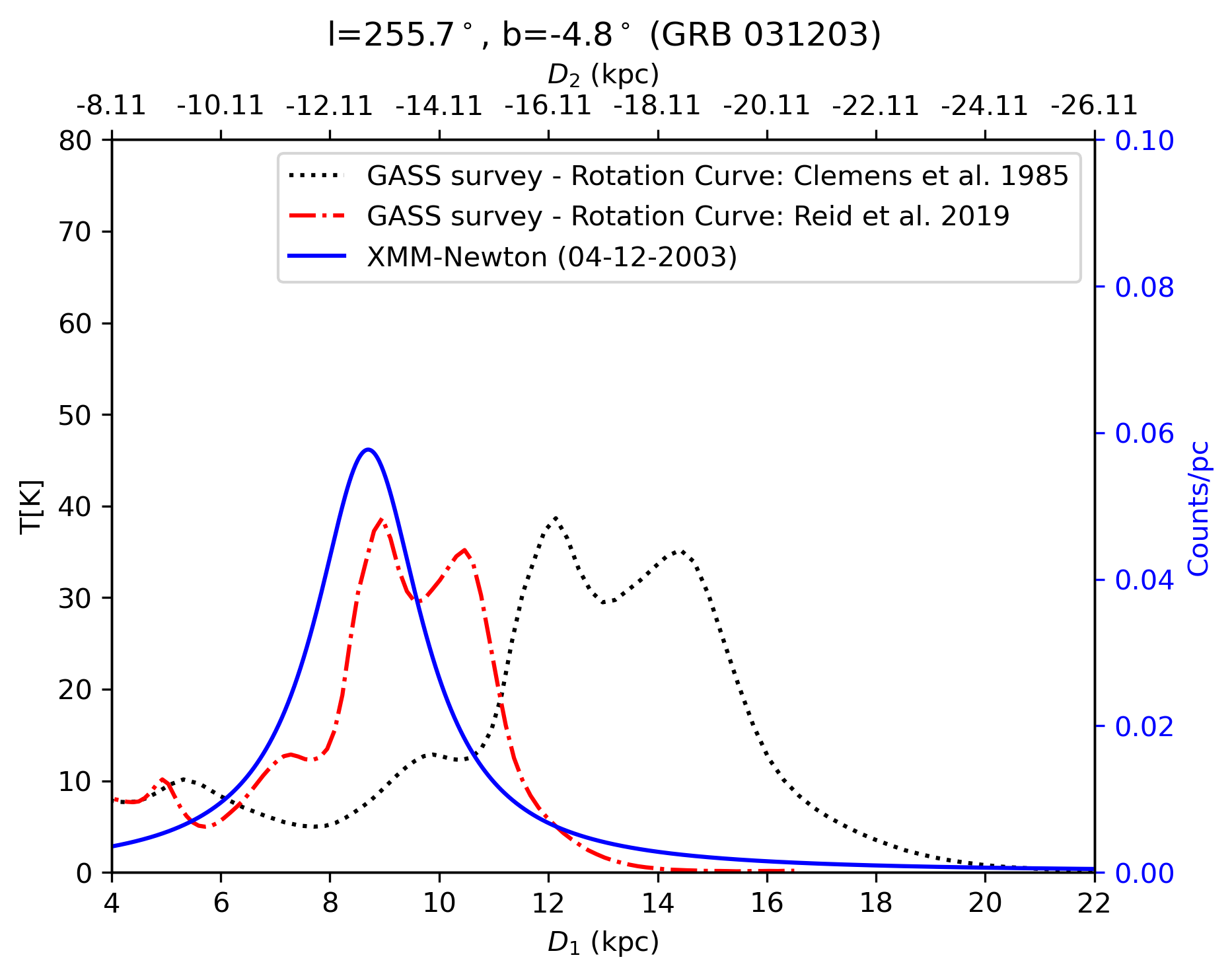}    
        \label{EBHIS3}
    \end{subfigure}
\caption{Comparison of our dust distance measurements (blue solid lines) in the directions of GRB 221009A (upper panel), GRB 160623A (central panel) and GRB 031203 (lower panel)  with \ion{H}{I} data from the EBHIS survey \citep{Winkel2016} for GRB221009A and GRB100623A, and from the GASS survey \citep{GASS} for GRB031203. The LSR velocities are deprojected into distances using two different models for the Galactic rotation curve: \citet{Reid2019} (red dot-dashed lines) and \citet{Clemens1985} (black dotted lines).}
\label{EBHIS}
\end{figure}

\ion{H}{I} and CO emission can trace the large-scale structure of spiral arms across the Galaxy, but their interpretation relies on Galactic kinematic models, introducing systematic uncertainties at large distances. Figure~\ref{EBHIS} presents a direct comparison of our distance measurements (blue solid lines) for the directions of GRB 221009A and GRB 160623A with \ion{H}{I} data from the EBHIS survey \citep{Winkel2016}, which covers the entire northern sky. For GRB 031203 direction, located in the southern hemisphere, a similar comparison is made using \ion{H}{I} data from the GASS survey \citep{GASS}. The \ion{H}{I} data are extracted from a $15^\prime$ circular region around the GRB directions for GRB 221009A and GRB 160623A, and a $30^\prime$ region for GRB 031203.

The local standard of rest velocities of the \ion{H}{I} clouds are deprojected into distances using two different Galactic rotation curve models: \citet{Reid2019}, which is the one used in  Fig.~\ref{Galaxy}, and \citet{Clemens1985}.

These two rotation curves are among the most widely used in the literature: \citet{Clemens1985} is based on CO observations and adopts $R_{0}=8.5$ kpc and $\Theta_{0}=220$ km s$^{-1}$, while \citet{Reid2019}, derived from maser parallaxes of high-mass star-forming regions, assumes $R_{0}=8.15$ kpc and $\Theta_{0}=236$ km s$^{-1}$. For each velocity, the deprojection yields two possible distances that are displayed along the two horizontal axes in Figure~\ref{EBHIS}.

In general, the Galactic rotation curve model of \citet{Reid2019} provides a better match to the locations of the dust clouds derived in this work compared to the \citet{Clemens1985} model. However, at the expected distances of the Outer and the OSC arms, even the \citet{Reid2019} model appears to diverge systematically from the distribution of the dust clouds, suggesting possible deviations in the assumed Galactic dynamics. In particular, in the directions of GRB 221009A and GRB 160623A, the peaks obtained from the \citet{Reid2019} model are systematically at smaller distances than those derived from X-ray data, whereas the \citet{Clemens1985} rotation curve always provides larger distances.

\begin{figure*}[h!]
\centering
\includegraphics[width=.7\textwidth]{image.png}
\caption{Spiral structure of the Milky Way, adapted from \citet{Sun2024}. The background map shows the distribution of molecular clouds (MCs) and high-mass star-forming regions (HMSFRs). Red and cyan circles indicate the 1306 MCs from \citet{Sun2024} and 10 MCs from \citet{Dame2011}, respectively. The white dashed lines represent the symmetrical Scutum - Centaurus and Perseus arms. The red square with error bar marks the only directly measured trigonometric parallax for the OSC arm \citep{Sanna2017}, while the diamond represents the kinematically derived location of the same HMSFR. Blue pluses and crosses indicate distant H II regions and star-forming masers, respectively.
Our measured distances are overlaid as cyan dots, showing the positions and the widths (surrounding red circles) of the spiral arms as determined from X-ray scattering rings of GRB 221009A (Rings 17, 18, and 19), GRB 160623A, and GRB 031203. The directions of these three GRBs are highlighted in purple.}
\label{Galaxy}
\end{figure*}

We also examined the sightlines toward the three GRBs using the CO survey by \citet{Dame2001}, extracting data in the GRB directions and averaging over a $1^\circ$ region. However, no significant CO emission was detected at distances greater than 5 kpc.

\subsection{Perseus Arm}

As shown in Fig. \ref{Galaxy}, the directions of the 3 GRBs cross the Perseus Arm. GRB 031203 is located very close to the end of this spiral arm, and no X-ray ring at the corresponding distance has been detected. 

This result is consistent with the lack of a prominent peak at the same distance in the GASS data (bottom panel of Fig.~\ref{EBHIS}), which confirms that a negligible amount of dust content is present in the Perseus Arm in the GRB direction ($l=256\degree$, $b= -5\degree$).

Dust scattering rings can instead be associated to the Perseus Arm in the X-ray observations of GRB 221009A and GRB 160623A, allowing us to accurately measure its distance towards their directions: 
$9.6\pm0.1$ and $5.12\pm0.03$ kpc, respectively.

Both these independent measurements are fully consistent with a symmetric inner-Galaxy spiral structure \citep{Vallee2008}, as shown by the white dashed lines in Fig. \ref{Galaxy}.

Also the EBHIS data in the direction of GRB 221009A (top panel of Fig.~\ref{EBHIS}) display a clear peak at $\sim$10 kpc, with a slightly better agreement with X-ray data if the \citet{Clemens1985} rotation curve is adopted. The EBHIS data toward GRB 160623A (middle panel of Fig.~\ref{EBHIS}) indicate a complex structure in the Perseus arm. In particular, we observe a broad bump at a distance consistent with that derived from the dust-scattering ring reported by \citet{Pintore2017}, as well as a smaller peak compatible with the new X-ray signal detected at 6.9 kpc.
The detection of significant amounts of dust at the Galactic longitudes of GRB 221009A ($l=53\degree$) and GRB 160623A ($l=84\degree$) is particularly interesting because they are close to the edges of the Perseus Arm Gap, which is a $\sim$6 kpc section of the spiral arm with a reduced massive star formation activity \citep{zhang13, sakai22}.

\subsection{Outer Arm}
The Outer Arm is the only spiral arm that could be mapped in all three directions. The detection of a significant excess in the pseudo-distance distributions derived from two different \textit{XMM-Newton} observations of GRB 031203 with distance and width consistent with those expected for the Outer Arm in this direction, allows us to confirm that this spiral arm extends up to $l = 256\degree$ and $b = -5\degree$. A clear signal with a double-peaked profile is detected in the GASS data (bottom panel of Fig.~\ref{EBHIS}). Adopting the \citet{Reid2019} rotation curve, the inferred distance is comparable to that derived from the X-ray analysis, whereas assuming the kinematic parameters of \citet{Clemens1985} yields a significantly larger value.

The distances measured for Ring 18 in GRB 221009A ($13.9\pm0.1$ kpc) and for the smallest newly discovered ring in GRB 160623A ($9.9\pm0.6$ kpc) exceed those predicted by the model in Fig. \ref{Galaxy}.
This suggests that the \citet{Reid2019} rotation curve used by \citet{Sun2024} to derive the Milky Way structure underestimates the distance of the Outer Arm in this direction. In contrast, the top and middle panels of Fig.~\ref{EBHIS} indicate that the opposite is valid for the \citet{Clemens1985} rotation curve.

\subsection{Outer Scutum-Centaurus Arm}
The OSC arm was clearly detected only in the direction of GRB 221009A, positioning the arm farther than both models shown in  Fig. \ref{Galaxy}: the dashed black line is the extrapolation of the symmetrical spiral structure by \citet{Vallee2008} and the solid black line is derived from the analysis of a selection of molecular clouds \citep{Sun2024}. This discrepancy is confirmed by the EBHIS data with distances derived using the same rotation curve \citep{Reid2019}, while an excessive distance is obtained assuming the \citet{Clemens1985} kinematic parameters (top panel of Fig.~\ref{EBHIS}). Since the same effect was observed also for the Outer Arm, our analysis suggests a revision of both rotation curves to better reproduce the structure of the two outer arms of our Galaxy in the First Quadrant.

The other distance measurements of individual objects reported in Fig. \ref{Galaxy} are spread over a relatively wide distance range and are affected by large uncertainties. The choice of a specific rotation curve (\citealt{Reid2019} in this case) leads to an underestimation of all distances derived from kinematic methods. On the other hand, the most robust distance measurement previously reported for this arm ($20.4^{+2.8}_{-2.2}$ kpc; \citealt{Sanna2017}), based on the parallax of a water maser source at small Galactic longitude (red square with error bar in Fig. \ref{Galaxy}), has an order of magnitude larger uncertainty than our measurement ($19.0\pm0.2$ kpc from \textit{Chandra} data). 

Moreover, the peak width of $\sim$2 kpc reflects the radial extent of dust along the line of sight, supporting the interpretation that it traces the arm's thickness rather than a single localized cloud.

The detection of a peak at 19 kpc in the pseudo-distance distribution might be surprising even for an exceptionally bright event as GRB 221009A, since its latitude of 4.3\degree implies the presence of a significant amount of dust 1.4 kpc above the Galactic plane. However, this portion of the OSC Arm is significantly inclined, with an average latitude of $\sim3\degree$ and a width $>2\degree$ at the longitude of GRB 221009A \citep{Dame2011}. Therefore, the non-detection of a similar structure in the GRB 160623A is justified not only by the lower fluence of this GRB, but also by its negative latitude ($b=-2.7\degree$).

\section{Conclusions}

Combining \textit{XMM-Newton} and \textit{Chandra} observations of three low-latitude GRBs in different directions, we could map multiple Galactic spiral arms, cross-check distances, and provide robust constraints on the large-scale geometry of the outer Galaxy.  In particular, our measurement of the OSC arm places it at 19.0 $\pm$ 0.2 kpc, achieving a precision far superior to the previous maser-parallax measurements ($20.4^{+2.8}_{-2.2}$ kpc; \citealt{Sanna2017}).

Although low-latitude GRBs as bright as GRB 221009A are very unlikely to repeat soon, future X-ray observatories such as \textit{AXIS} \citep{AXIScite} and \textit{New Athena} \citep{NewAthenaCite} will enable systematic campaigns of GRB follow-ups, which, thanks to their higher sensitivity, will expand the sample of X-ray rings generated by dust in the outskirts of the Milky Way. 

This will allow precise mapping of interstellar clouds traced by dust along more sightlines, complementing the 3D dust distributions derived from extinction studies in the optical and infrared band, based, for example, on future Gaia Data Releases and the Galactic plane Survey that will be performed by the \textit{Nancy Grace Roman Space Telescope} \citep{RomanCite}.

\begin{acknowledgements}
This publication was produced while B.V. attending the PhD program in Space Science and Technology at the University of Trento, Cycle XXXVIII, with the support of a scholarship financed by the Ministerial Decree no. 351 of 9th April 2022, based on the NRRP - funded by the European Union - NextGenerationEU - Mission 4 "Education and Research", Component 1 "Enhancement of the offer of educational services: from nurseries to universities" - Investment 4.1 "Extension of the number of research doctorates and innovative doctorates for public administration and cultural heritage". This paper and related research have been conducted during and with the support of the Italian national inter-university PhD programme in Space Science and Technology - University of Trento / IUSS Pavia. AB acknowledges the financial support of INAF through the initiative "IAF Astronomy Fellowships in Italy" (grant name MEGASKAT).

\end{acknowledgements}

\bibliography{main}
\bibliographystyle{aa}

\end{document}